\documentclass[twocolumn,showpacs,amsmath,pra,aps,amssymb,superscriptaddress,nofootinbib]{revtex4-1} 
\usepackage[T1]{fontenc}
\usepackage[latin9]{inputenc}
\usepackage{times}
\usepackage{color} 
\usepackage{xspace}
\usepackage{amssymb,amsmath}
\usepackage{amsbsy}
\usepackage[pdftex]{graphicx}
\usepackage{bm}
\usepackage{float}
\usepackage{adjustbox}
\usepackage{epstopdf}

\usepackage[unicode,breaklinks]{hyperref}
\hypersetup{
    unicode=true,
    plainpages=false, 
    colorlinks=true,
    linkcolor=blue,
    citecolor=blue,
    filecolor=black,
    urlcolor=blue
}
\usepackage{url}
\usepackage{verbatim}

\begin{document}

\title{Nematic ordering dynamics of an anti-ferromagnetic spin-1 condensate}
\author{L.~M.~Symes}  
\affiliation{Dodd-Walls Centre for Photonic and Quantum Technologies, Department of Physics, University of Otago, Dunedin 9016, New Zealand}
\author{P.~B.~Blakie}  
\affiliation{Dodd-Walls Centre for Photonic and Quantum Technologies, Department of Physics, University of Otago, Dunedin 9016, New Zealand}
\affiliation{Swinburne University of Technology, Sarawak Campus, School of Engineering, Computing and Science,  Jalan Simpang Tiga, 93350 Kuching, Sarawak, Malaysia}

\begin{abstract}
We consider the formation of order in a quasi-two-dimensional (quasi-2D) anti-ferromagnetic spin-1 condensate quenched from an easy-axis (EA) to an easy-plane (EP) nematic phase. We define the relevant order parameter to quantify the spin-nematic degrees of freedom and study the evolution of the spin-nematic and superfluid order during the coarsening dynamics using numerical simulations. We observe dynamical scaling in the late time dynamics with both types of order extending across the system with a diffusive growth law. We identify half-quantum vortices (HQVs) as the relevant topological defects of the ordering dynamics, and demonstrate that the growth of both types of order is determined by the mutual annihilation of these vortices.
\end{abstract}

\maketitle

\section{Introduction}
Spin-1 condensates \cite{Stenger1998a,Ho1998a,Ohmi1998a} with anti-ferromagnetic interactions prefer to order into spin-nematic phases \cite{Kawaguchi2012R}. Such phases have a vanishing average spin-density, and are instead characterized by the  nematic tensor $\mathcal{N}_{ab}=\tfrac{1}{2}\langle f_af_b+f_bf_a\rangle$, where $f_{a\in\{x,y,z\}}$ are the spin matrices.  The ground states of this system have an axially symmetric nematic tensor (uniaxial nematic) with a preferred axis (but not direction) characterized by a director $\vec{u}$ in spin-space (i.e.~$\vec{u}$ and $-\vec{u}$ are equivalent). Recently   experimental evidence was presented for spin-nematic order  in an anti-ferromagnetic condensate \cite{Zibold2016a}.  

The concept of nematic order is typically discussed in the context of liquid crystals, where the order is associated with the orientation of long molecules. Indeed, many beautiful studies of phase transition dynamics and coarsening have been performed in liquid crystal systems (e.g. see \cite{Orihara1986a,Chuang1991a,Chuang1991b,Yurke1992a,Nagaya1992a,Blundell1992a,Chuang1993a}). A sudden change in conditions (e.g.~temperature or pressure) is used to quench this system from an isotropic phase (unoriented molecules) into the nematic phase, and the formation of order and defect dynamics can be observed optically.

\begin{figure}
\vspace*{-0.15cm}
\includegraphics[width=2.8in]{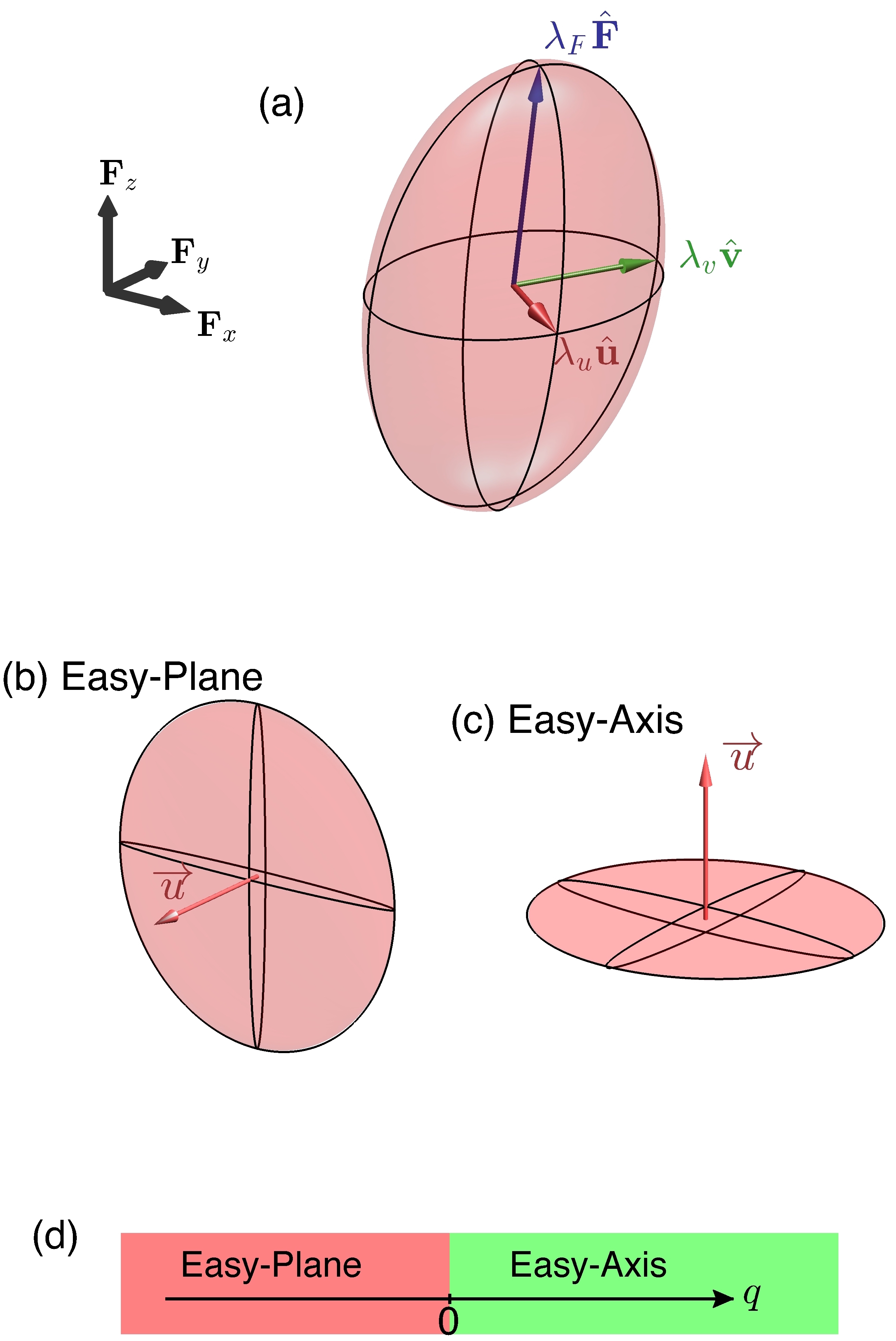}
\vspace*{-0.175cm}
\caption{\label{nematicFig}  
(a) Representation of the nematic tensor $\mathcal{N}$ of a spin-1 state as an ellipsoid. The semi-principal axes are along  eigenvectors of $\mathcal{N}$ indicated by the unit vectors of $\{\vec{u},\vec{v},\mathbf{F}\}$, with the widths in these directions given by corresponding eigenvalues $\{\lambda_u,\lambda_v,\lambda_F\}$. A polar state is a flat disk-shaped ellipsoid [see (b) and (c)] completely characterized by the director $\vec{u}$, with spin fluctuations maximized in the plane transverse to $\vec{u}$. (b) Easy-plane  and (c) Easy-axis cases of the polar state.    (d) Ground state phase diagram as a function of $q$. Note that the direction of the external magnetic field sets our $z$-axis. }
\end{figure}

In this paper we develop a theory for the ordering dynamics (coarsening) of an anti-ferromagnetic spin-1 condensate.  There has been considerable theoretical work on the coarsening dynamics of ferromagnetic spin-1 condensates \cite{Mukerjee2007a,Kudo2013a,Kudo2015a,Williamson2016a}, however this area is largely unexplored in the anti-ferromagnetic system. 
Our interest is in the symmetry breaking phase transition from an EA phase (with $\vec{u}$ along the direction set by the external magnetic field) at positive quadratic Zeeman energy $q$, to an EP phase ($\vec{u}$ transverse to the external field) at negative $q$ (see Fig.~\ref{nematicFig}(d) and Refs.~\cite{Bookjans2011b,Vinit2013a,NguyenThanh2013a,Jiang2014a,Vinit2016a}).
We consider a quench between these phases implemented by a sudden change in $q$, e.g.~using microwave dressing (see \cite{Bookjans2011b,Vinit2013a,Jiang2014a}).  Upon entering this new phase, the system breaks the continuous axial symmetry of the initial state by developing transverse spin-nematic domains. Here our interest lies in characterizing the dynamics of the phase transition, with a particular emphasize on the late-time coarsening dynamics. That is, to understand the universal aspects of how small domains created after the quench anneal together to bring the system towards an ordered equilibrium state.
To undertake this study we first discuss how the nematic order is characterized in a spinor condensate, and develop an appropriate order parameter for the EP phase. Using numerical simulations we study how the EP order forms in the system. We demonstrate that the late-time coarsening behaviour exhibits dynamical scaling with a diffusive domain growth law of $L(t)\sim [t/\ln(t)]^{1/2}$, where $L$ is the size of the ordered domains and $t$ is the time after the quench. We separately consider the superfluid order  and show that it grows with an identical law to the spin-nematic order, in contrast to recent results for the ferromagnetic spin-1 system \cite{Andreane2017a}.
The order parameter growth is determined by the dynamics of HQVs in the system, and we verify that the number of these vortices scales as $L(t)^{-2}$, i.e.~that coarsening proceeds by vortex anti-vortex pairs mutually annihilating.  Recent experiments have demonstrated that it is possible to measure HQVs in anti-ferromagnetic spin-1 condensates \cite{Seo2015a,Seo2016a} due to their ferromagnetic cores \cite{Lovegrove2012a}. Thus, measuring the HQV distribution as a function of time after the quench could be a practical method for experiments to quantify the coarsening of this system. Alternatively, it may be possible to directly image  \cite{Carusotto2004a,Higbie2005a} or probe \cite{Baillie2016a} nematic properties of the condensate.

We note that the symmetries and defects of the EP phase are similar to those of a (two-component) binary condensate in the miscible regime. Indeed, work by Karl \textit{et al.}~\cite{Karl2013a} discussed the role of the equivalent vortices in the ordering dynamics of a two-component system, although that work focused on understanding the emergence of power-law behavior in various momentum correlation functions, and relating these to turbulence cascades.

The outline of the paper is as follows. In Sec.~\ref{SecFormalism} we introduce the basic formalism for spin-1 condensates and consider how to quantify spin-nematic order. We discuss the EA to EP quench and introduce the relevant order parameter for this phase transition.  In Sec.~\ref{SecResults} we start by introducing the quasi-2D system,  the equation of motion and simulation technique we use to study the quench dynamics. We present results for the evolution of various local densities and correlation functions that illustrate the early time dynamics of the quench, and show the emergence of EP order. We then focus on the late time dynamics of the system and characterise the phase ordering dynamics. To do this we introduce correlation functions for the spin-nematic and superfluid order. We evaluate these using an  ensemble of large-scale simulations and demonstrate correlation function collapse (dynamic scaling) and extract the relevant growth laws. Finally, we examine the role of  HQVs and show that the average distance between vortices characterises the growth of order. Then we conclude in Sec.~\ref{Sec:Conclude}.

\section{Formalism}\label{SecFormalism}
\subsection{Spin-1 Anti-ferromagnetic condensate}
A spin-1 condensate is described by the spinor field
\begin{align}
\bm{\psi}\equiv (\psi_{1},\psi_0,\psi_{-1})^T,\label{Eqpsisph}
\end{align} 
where the three components describe the condensate amplitude in the spin levels $m=1,0,-1$, respectively. 
The short-ranged interactions between atoms are described by the rotationally invariant Hamiltonian density
\begin{align}\label{spinH}
\mathcal{H}_{\mathrm{int}}= \frac{g_n}{2}n^2+\frac{g_s}{2}\left|\bm{F}\right|^2.
\end{align}
The first term describes the density dependent interactions, with coupling constant $g_n$, where  $n\equiv\bm{\psi}^\dagger\bm{\psi}$ is the total density.  The second term describes the  spin-dependent interactions $g_s|\bm{F}|^2$,  with coupling constant $g_s$,  where  $\bm{F}\equiv \bm{\psi}^\dagger\bm{f}\bm{\psi}$  is the spin density and  $\bm{f} \equiv (f_x,f_y,f_z)$ are the  spin-1 matrices. For the case $g_s>0$, known as anti-ferromagnetic interactions, the condensate prefers to minimise the spin-density to reduce the interaction energy.
In addition to interactions, the (uniform) quadratic Zeeman shift
\begin{align}
\mathcal{H}_{\mathrm{QZ}}=q\bm{\psi}^\dagger f_z^2\bm{\psi},
\end{align}
also plays a role in determining the preferred spin-ordering of the condensate.
The quadratic Zeeman energy $q$ can be controlled using the magnetic bias field, it can also be varied by using microwave dressing (e.g.~see \cite{Gerbier2006a,Leslie2009a}).

\subsection{Nematic order}\label{SecPDSymmetries}
To quantify the spin-order it is useful to introduce the Cartesian representation of the spinor field $\vec{\psi}\equiv(\psi_x,\psi_y,\psi_z)$, where $\psi_x=(\psi_{-1}-\psi_{1})/\sqrt{2}$,   $\psi_y=-i(\psi_1+\psi_{-1})/\sqrt{2}$, and $\psi_z=\psi_0$. We will give results in both the cartesian $\vec{\psi}$ and spherical [$\bm{\psi}$, see Eq.~(\ref{Eqpsisph})] bases as needed.

A general spinor can be decomposed in the form
\begin{align}
\vec{\psi}=e^{i\theta}(\vec{u}+i\vec{v}),\label{EqDecomp}
\end{align}
where $\theta$ is the global phase and $\{\vec{u},\vec{v}\}$ are mutually orthogonal real vectors satisfying $|\vec{u}|^2+|\vec{v}|^2=n$, and $|\vec{u}|\ge|\vec{v}|$ (also see \cite{Ohmi1998a,Mueller2004a,Yukawa2012,Zibold2016a}). 
For a spin-1 spinor, the local spin information described by the spin density vector is
\begin{align}
\mathbf{F}=-i\vec{\psi}^*\times\vec{\psi}=2\vec{u}\times\vec{v},
\end{align}
and the symmetric nematic (or quadrupolar) tensor density is
\begin{align}
\mathcal{N}_{ab}&=\frac{1}{2}\langle f_af_b+f_bf_a\rangle,\qquad a,b \in \{x,y,z\}\\
&=n\delta_{ab}-\frac{1}{2}(\vec{\psi}^*\otimes\vec{\psi}+\vec{\psi}\otimes\vec{\psi}^*), \\
&=n\delta_{ab}-(\vec{u}\otimes\vec{u}+\vec{v}\otimes\vec{v}).
\end{align}
The nematic tensor describes the anisotropy of the spin fluctuations, and in general has the symmetries of an ellipsoid. This is revealed by diagonalizing $\mathcal{N}$, giving $\{\vec{u},\vec{v},\mathbf{F}\}$ as the eigenvectors with respective eigenvalues $\lambda_u=\frac{1}{2}(n-\mathcal{A})$, $\lambda_v=\frac{1}{2}(n+\mathcal{A})$ and $\lambda_F=n$. 
 Here $\mathcal{A}=2|\vec{u}|^2-n\ge0$ is the \textit{alignment parameter} \cite{Zibold2016a}, which characterizes the relative fluctuations of magnetization  along the directions orthogonal to $\mathbf{F}$.  
The alignment is related to the spin-singlet amplitude\footnote{Note our definition differs by a constant factor from \cite{Kawaguchi2012R}.} \begin{align}
\alpha=\vec{\psi}\cdot\vec{\psi}=\psi_0^2-2\psi_{1}\psi_{-1},\label{Alpha}
\end{align}
 as $\mathcal{A}=|\alpha|$. It is conventional to take the eigenvector associated with the smallest eigenvalue of $\mathcal{N}$  as the nematic director, i.e. the vector $\vec{u}$. We can use the eigenvectors and eigenvalues to represent the nematic tensor density as an ellipsoid [see Fig.~\ref{nematicFig}(a)].
 We also note that $\lambda_u=|\vec{v}|^2$ and $\lambda_v=|\vec{u}|^2$, so that the extent of the ellipsoid along the $\vec{u}$ direction is the squared length of $\vec{v}$, and the extent of the ellipsoid   along the $\vec{v}$ direction is the squared length of $\vec{u}$.

Two limiting states are of interest. First, the fully magnetized ferromagnetic state with $|\mathbf{F}|=n$, where $|\vec{u}|=|\vec{v}|=\sqrt{n/2}$, and   $\mathcal{A}=0$. Second, and of primary concern in our work, is the fully polar (or spin-nematic) state which has the form
\begin{align}
\vec{\psi}_\mathrm{P}=e^{i\theta}\vec{u},\label{PsiP}
\end{align}
with $|\vec{u}|=\sqrt{n}$, $\mathcal{A}=n$ and $\mathbf{F}=0$ [see Figs.~\ref{nematicFig}(b) and (c)].  The spin properties of this state are completely characterized by the director $\vec{u}$, and the state is invariant under the transformation 
\begin{align}
\theta\to\theta+\pi, \quad\mathrm{and}\quad \vec{u}\to-\vec{u}.\label{EqDiscretesymm}
\end{align}
For general spin-1 states the relation 
\begin{align}
|\mathbf{F}|^2+\mathcal{A}^2=n^2,\label{EqAlignReln}
\end{align}
holds, so that  $\mathcal{A}$ can be used to characterize how close a state is to the limiting cases of ferromagnetic ($\mathcal{A}=0$) or polar ($\mathcal{A}=n$) order.

\subsection{Order parameter for the EA to EP phase transition}\label{SecOrderEAEP}
Here we are concerned with an anti-ferromagnetic condensate in which a quench is performed by a sudden change in the quadratic Zeeman energy from a positive value to a negative value\footnote{The $z$-magnetization $M_z\equiv\int d^2\mathbf{x}\,F_z$ of the system is conserved, and here we focus on the case $M_z=0$ where the transition occurs at $q=0$. } crossing a quantum phase transition between two different ground states [see Fig.~\ref{nematicFig}(d)].  For both cases the ground state is fully polar $\vec{\psi}=e^{i\theta}\vec{u}$. For $q>0$ the director ($\vec{u}$) is along the $z$ axis [EA phase, see Fig.~\ref{nematicFig}(c)]. For $q<0$ the director lies in the $xy$-plane [EP phase, see Fig.~\ref{nematicFig}(b)]. Thus the EP phase breaks the axial symmetry (invariance to spin rotations about $z$) of the Hamiltonian.
 This type of quench in an anti-ferromagnetic spinor condensate of $^{23}$Na atoms has been performed in a number of experiments \cite{Bookjans2011b,Vinit2013a,Jiang2014a,Vinit2016a,Kang2017a}, however the EP nematic order was not directly probed in these studies  (c.f.~\cite{Zibold2016a}). 
We also note that other phase transitions can be considered in this system, e.g.~Witkowska \textit{et al.}~\cite{Witkowska2014a} considered a $q$ quench for an anti-ferromagnetic condensate with a non-zero (conserved) $z$-magnetization, where a transition to a phase separated state occurs.

We would like to obtain an order parameter that can distinguish between these two states, notably the order parameter should be zero in the EA phase and non-zero in the EP phase. 
To do this we note that in the EA phase the nematic tensor is isotropic in the $xy$-plane [see Fig.~\ref{nematicFig}(c)], while in the EP phase the nematic tensor is anisotropic in the $xy$-plane [see Fig.~\ref{nematicFig}(b)]. To quantify the EP nematic order, and taking motivation from nematic liquid crystals \cite{DeGennesBook}, we use a traceless symmetric tensor to quantify order in this system. Particular to the EA to EP phase transition we use the planar tensor:
\begin{align}
Q &= \mathcal{N}_{2 \times 2} - \tfrac{1}{2} \mathrm{Tr}\{\mathcal{N}_{2 \times 2}\} I_2, \label{EqQ}\\
&= \left(\begin{array}{cc}Q_{xx} & Q_{xy}\\ Q_{xy} & -Q_{xx}\end{array}\right),
\end{align}
where $\mathcal{N}_{2 \times 2}$ is the $xy$-submatrix of $\mathcal{N}$, and $I_2$ is the identity matrix. Evaluating this expression we find that $Q_{xx}=\mathrm{Re}\{\psi_1^*\psi_{-1}\}$ and $Q_{xy}=\mathrm{Im}\{\psi_1^*\psi_{-1}\}$, i.e.~it depends on the relative phase coherence between the $\psi_1$ and $\psi_{-1}$ components of the system. 
While $Q$ is traceless by construction, $\mathrm{Tr}(Q^2)=0$ only when the spin fluctuations are isotropic in the $xy$-plane. The EP phase is thus revealed by $\mathrm{Tr}(Q^2)$  
becoming non-zero, thus demonstrating how $Q$ serves as an order parameter.  We can write the eigenvalues of $Q$ as $\{-\frac{1}{2}\mathcal{A}_\perp,\frac{1}{2}\mathcal{A}_\perp\}$, where we have defined a ``transverse alignment'' parameter\footnote{$\mathcal{A}_\perp$ is sensitive to anisotropy of $ {Q}$, but does not completely distinguish between polar and ferromagnetic states as does $\mathcal{A}$. E.g., the fully ferromagnetic state with $\mathbf{F}=n\hat{\mathbf{x}}$ has $\mathcal{A}=0$, but $\mathcal{A}_\perp=\frac{1}{2}n$ (c.f.~the pure EP polar state with $\vec{u}=\sqrt{n}\hat{\mathbf{x}}$ for which $\mathcal{A}=\mathcal{A}_\perp=n$). As is apparent from Fig.~\ref{nematicFig}(a) a state with $\mathbf{F}\ne0$ also has an anisotropic nematic tensor, just to a lesser extent than a polar state.} 
\begin{align}
\mathcal{A}_\perp=|\alpha_\perp|,\label{EqAperp}
\end{align}
and have introduced [c.f.~Eq.~(\ref{Alpha})]
 \begin{align}
\alpha_\perp \equiv-2\psi_{1}\psi_{-1}.\label{AlphaPerp}
 \end{align}
 Using this result gives $\mathrm{Tr}(Q^2)=\frac{1}{2}\mathcal{A}_\perp^2$.  In Appendix \ref{AppPlanar} we present an alternative formulation of the planar tensor $Q$ and order parameter results.

\section{Results}\label{SecResults}
\subsection{Quasi-two-dimensional quench}
In order to explore the quench dynamics we focus on a  quasi-2D  system. In this regime the extent of the condensate in one direction (which we take to be $z$) is less than the spin healing length, so spin motion is effectively frozen out in this direction. This regime has been realized in experiments by applying a tight optical trap in this direction (e.g.~see \cite{Sadler2006a,Seo2015a}). Additionally, we neglect any transverse confinement and take the condensate to be homogeneous in the plane. The dynamics of this system is described by  the spin-1 Gross-Pitaevskii equation (GPE)
\begin{align}\label{spinGPE}
i\hbar\frac{\partial\bm{\psi}}{\partial t}=\left(-\frac{\hbar^2\nabla^2}{2M}+qf_z^2+g_nn+g_s\bm{F}\cdot\bm{f}\right)\bm{\psi}.
\end{align} 
Note we have neglected the linear Zeeman shift which can be removed from the equation of motion by transforming to a rotating frame.

To numerically solve this equation we represent each component of the spinor field $\bm{\psi}$ on a 2D square region of dimensions $l\times l$ covered by an $N\times N$ grid of equally spaced points. Taking periodic boundary conditions for the solution we evaluate spatial derivatives in the kinetic energy term of Eq.~(\ref{spinGPE}) with spectral accuracy using fast Fourier transforms. To evolve the GPE in time we use the second order symplectic method presented in Ref.~\cite{Symes2016a}. 

The initial condition for the simulations is a uniform EA ground state (in the spherical basis)
\begin{align}
\bm{\psi}(\mathbf{x},t=0)=\sqrt{n_c}\left(\begin{array}{c}0\\1\\0\end{array}\right)+\bm{\delta}(\mathbf{x}),
\end{align}
where ${n_c}$ is the condensate (areal) density and $\bm{\delta}$ is a small noise field added to seed the growth of unstable modes following the quench. The late-time results are insensitive to the form of white spatial noise we add to the initial state as long as the noise is weak ($|\bm{\delta}|^2\ll n_c$). We choose to add noise according to the truncated Wigner prescription \cite{cfieldRev2008}, which is consistent with the quantum vacuum noise on the initial state (see \cite{Williamson2016b} for details).
We introduce the characteristic spin energy $q_0\equiv2g_sn_c$, and associated spin healing length $\xi_s=\hbar/\sqrt{Mq_0}$ and spin time $t_s=\hbar/q_0$ as convenient units.
 
\subsection{Early-time dynamics: development of local order}\label{SecEarlyTime}
\begin{figure}
\includegraphics[width=3.4in]{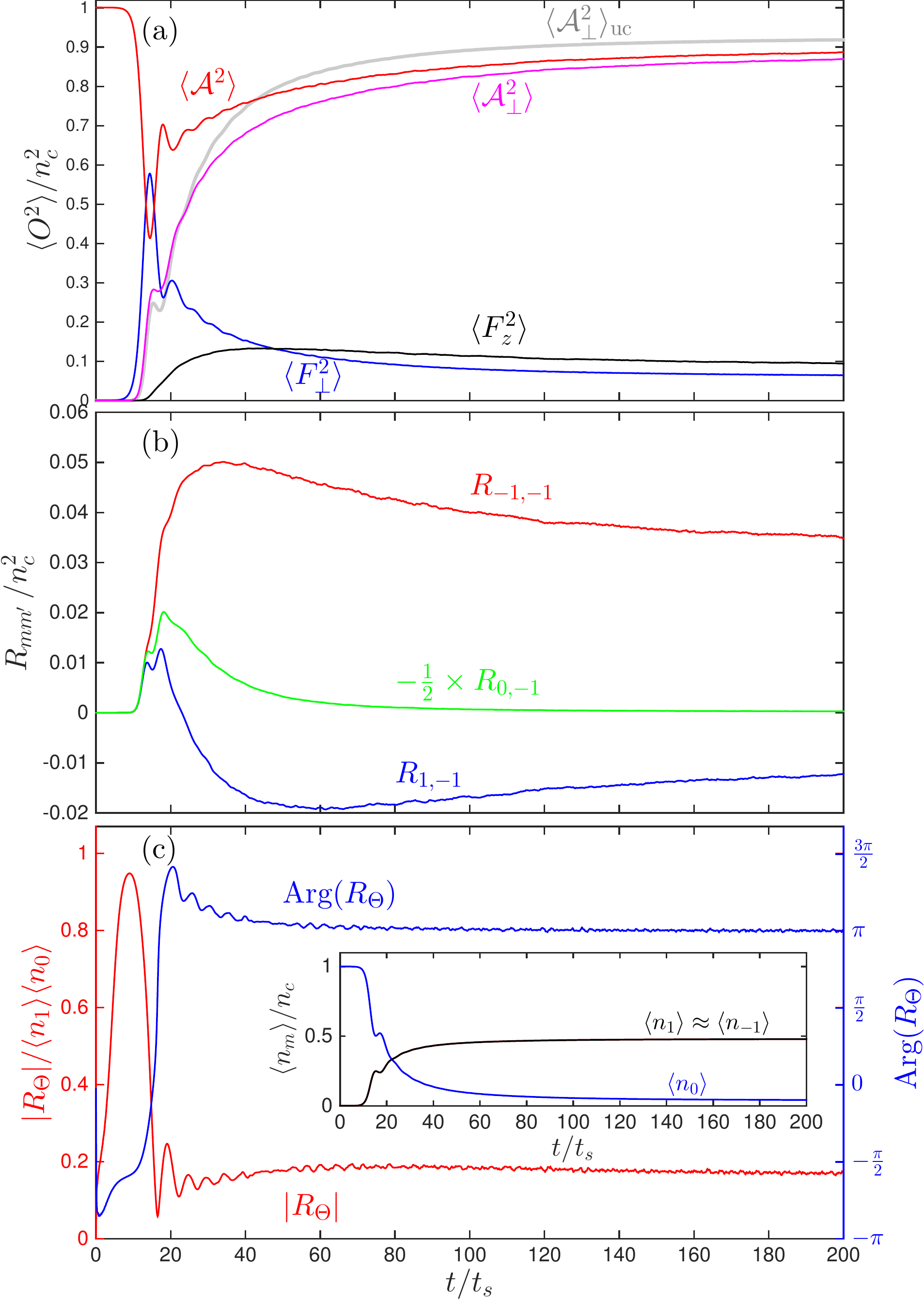} 
\caption{\label{LocalOrderFig} Growth of densities, local pair correlations  and $R_\Theta$ following a quench from the EA to EP phase. (a) The local densities $O=\{n, \mathcal{A},\mathcal{A}_\perp,\mathbf{F}\}$ are evaluated from the results of a single simulation trajectory according to Eq.~(\ref{Eqlocalorder}). (b) The local pair correlations functions, as defined in Eq.~(\ref{EqRmmp}). (c) The relative phase correlation function $R_\Theta$ as defined in the Eq.~(\ref{GTheta}). Inset: The evolutions of the mean component densities, noting that the $m=\pm1$ results are approximately identical.
Simulation is for a quench to $q=-0.5q_0$ with $g_n=3g_s$. The simulation is for a condensate density $n_c=10^4/\xi_s^2$ of size $l=400\,\xi_s$ with $N=512$ points in each direction.  }
\end{figure}

Immediately following the quench the initial EA state is unstable and begins to evolve towards the new phase. 
Aspects of these early time dynamics, and the emergence of local EP order can be revealed by studying the behaviour of the spin and alignment densities.
Since some of these densities (e.g.~$F_z$) can be locally negative, we quantify the development of a particular density of interest $O$ by spatially averaging  $O^2$, i.e.~we evaluate
\begin{align}
\langle O^2(t)\rangle= \frac{1}{l^2}\int d^2\mathbf{x}\,O^2(\mathbf{x},t).\label{Eqlocalorder}
\end{align}
We present results for a variety of densities of interest in Fig.~\ref{LocalOrderFig}(a).
These results show that  immediately following the quench the EA state becomes dynamically unstable to magnon excitations which grow exponentially and cause the system to develop transverse magnetization [i.e.~$\mathbf{F}_\perp=(F_x,F_y)$]. The precise nature of the instability and the wavevectors of the unstable modes depends upon the value of $q$, and aspects of this have already been explored in experiments \cite{Bookjans2011b,Vinit2013a,Vinit2016a,Kang2017a}. The axial magnetization ($F_z$) similarly experiences exponential growth. The general behavior of  spin density growth we observe is similar for quenched condensates with ferromagnetic interactions (e.g.~see \cite{Saito2007a,Lamacraft2007a,Leslie2009a,Barnett2011}). Noting that the average $z$-magnetization of the initial state is zero (and conserved), the quantity $\langle F_z^2(t)\rangle$ corresponds to the fluctuations in magnetization  studied in recent experiments \cite{Kang2017a}.

More direct insight into the change in nematic order is provided by the alignment densities $\{\mathcal{A},\mathcal{A}_\perp\}$ discussed in Sec.~\ref{SecPDSymmetries}. The initial EA state is fully aligned (i.e.~$\mathcal{A}=n_c$), but this dips down in the early dynamics as the magnetization develops [as required by the relation (\ref{EqAlignReln})]. As the alignment is restored for $t\gtrsim20\,t_s$ it is of a different character, consistent with EP order emerging. We see this by evaluating the transverse alignment $\mathcal{A}_\perp$ order which is initially negligible, but  then grows and is seen to saturate towards the value of $\mathcal{A}$.

Various \textit{in situ} measurements of correlations between components of the density have been performed in spinor condensate experiments (e.g.~see \cite{Vinit2013a,Guzman2011a,Seo2016a,Kang2017a}). Most relevant to our system are the measurements of Vinit \textit{et al}.~\cite{Vinit2013a} of
the time evolution of the local pair correlation function following the EA to EP quench of a quasi-one-dimensional anti-ferromagnetic condensate. 
The correlation functions measured were\footnote{Here and for the remainder of this subsection all expectations will be taken to be spatially averaged as in Eq.~(\ref{Eqlocalorder}).}
\begin{align}
R_{mm'}(t)= \langle\delta n_m \delta n_{m'} \rangle,\label{EqRmmp}
\end{align}
where $\delta n_m(\mathbf{x},t)=n_m(\mathbf{x},t)-\langle n_m\rangle$ is the $m$-component density fluctuation operator, with $n_m=|\psi_m|^2$ and $\langle n_m\rangle$ being the mean density of this component.  We have evaluated the same correlation functions measured in experiments (c.f.~Fig.~3 of Ref.~\cite{Vinit2013a}) and present the results in Fig.~\ref{LocalOrderFig}(b). We find similar qualitative behavior to their results, however note that their measurements were for a shallow quench (to $q\approx-0.02q_0$) and with appreciable thermal effects.
These same types of local density measurements could be used to evaluate the alignment densities.  Indeed, noting that $\langle \mathcal{A}_\perp^2\rangle=4\langle n_1n_{-1}\rangle$ [see Eqs.~(\ref{EqAperp}) and (\ref{AlphaPerp})] , taking $ n_1$ and $ n_{-1} $ as uncorrelated, we can make the estimate
\begin{align}
\langle\mathcal{A}_\perp^2\rangle_{\mathrm{uc}} \approx {4\langle n_1\rangle\langle n_{-1}\rangle}.\label{EqAperpuc} 
\end{align}
For the uniform system   $\langle n_m\rangle=N_m/l^2$, and thus $\langle\mathcal{A}_\perp^2\rangle_{\mathrm{uc}}$  is determined by the component populations $N_m=\int d^2\mathbf{x}\,n_m$, which are readily measured in experiments.   As can be seen from Fig.~\ref{LocalOrderFig}(a) the uncorrelated approximation tends to overestimate the EP order ($\langle\mathcal{A}_\perp^2\rangle$) once it develops ($t\gtrsim20\,t_s$). Noting that $\langle n_1n_{-1}\rangle=\langle n_1\rangle\langle n_{-1}\rangle+R_{1,-1}$, this overestimate of Eq.~(\ref{EqAperpuc}) is due to the negative value $R_{1,-1}$ takes for $t\gtrsim20\,t_s$ [Fig.~\ref{LocalOrderFig}(b)]. Evidence for $R_{1,-1}$ becoming negative was also found in experiments at late times \cite{Vinit2013a}.

Finally we examine the system evolution to quantify the local ``phase locking'' of the $m=\pm1$ components relative to the $m=0$ component. This was recently observed in experiments by applying a spin rotation to the system and measuring the resulting magnetic fluctuations \cite{Zibold2016a}.
 In our simulations we can directly access this from the local (spatially averaged) correlation function
\begin{align}
R_\Theta(t) \equiv \left\langle \psi_{-1} \psi_{1} \psi_0^* \psi_0^* \right\rangle.\label{GTheta}
\end{align}
Taking $\psi_m=\sqrt{n_m}e^{i\theta_m}$, we see that $R_\Theta \sim e^{i(\theta_1+\theta_{-1}-2\theta_0)}$, which conventionally defines the relative phase $\Theta\equiv\theta_1+\theta_{-1}-2\theta_0$. To understand the physical relevance of  this correlation function, we note that the transverse spin density squared and the alignment density squared are
\begin{align}
\langle |F_{\perp}|^2 \rangle&=2\langle n_0(n_{-1} + n_{1})\rangle+4\mathrm{Re}\{R_\Theta\}, \label{FpT}\\
\langle \mathcal{A}^2 \rangle &= \langle n_0^2 \rangle + 4 \langle n_{1} n_{-1} \rangle - 4 \mathrm{Re}\{R_\Theta\},\label{AT}
\end{align}
respectively. 
Thus varying the real part of $R_\Theta$ the system can enhance or reduce the spin density, while having the opposite effect on the alignment [also see Eq.~(\ref{EqAlignReln})].
Anti-ferromagnetic systems prefer $\Theta=\pi$ to reduce the spin-density.  
The behaviour of $R_\Theta$ is shown in Fig.~\ref{LocalOrderFig}(c), noting that 
 we have normalized $R_\Theta$ by the average densities of each component [using $\langle n_{-1}\rangle\approx\langle n_1\rangle$, also see inset to Fig.~\ref{LocalOrderFig}(c)] so that the magnitude measures the concentration of $\Theta$.  These results show that after the early dynamics settles down ($t\gtrsim25t_s$) the function $R_\Theta$ approaches a negative real value, i.e.~$\Theta\to\pi$.
 The $m=0$ component is unoccupied in the EP ground state, but maintains a small population [see inset to Fig.~\ref{LocalOrderFig}(c)] at late times due to heating from the quench.  
The $m=0$ component  of the system is noisy (consistent with a thermalized gas, e.g.~see \cite{Williamson2016b}) and the amplitude of the $R_\Theta$ correlation function is significantly reduced by these fluctuations. However, our results show that there is still a tendency for the spin-dependent interactions to lock the relative phase of the $m=\pm1$ components relative to the $m=0$ component.

\begin{figure}[htb!]
\includegraphics[width=3.3in]{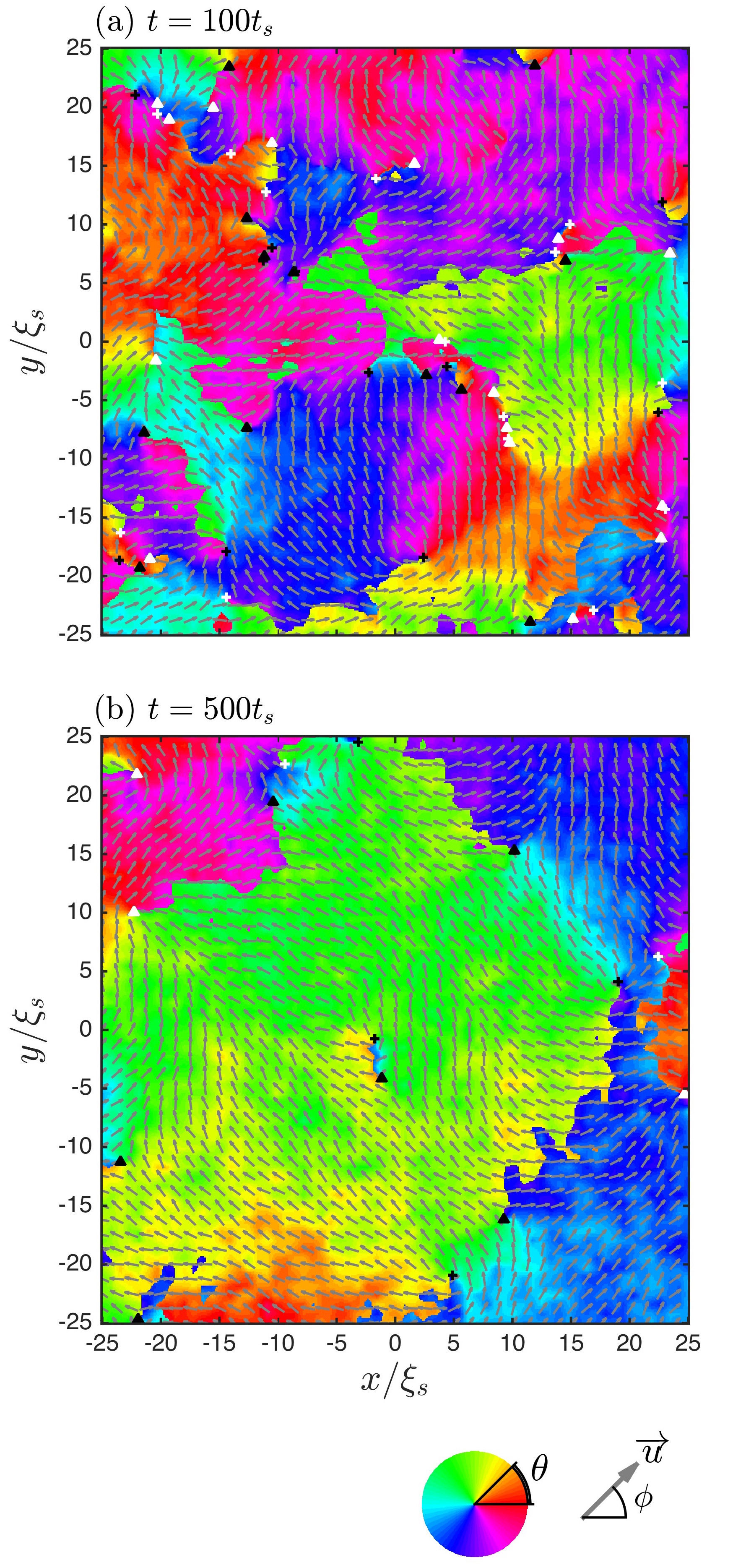}
\caption{\label{microstatesfig} Evolution of order after the quench in a $50\,\xi_s\times50\,\xi_s$ sub-region of a simulation at (a) $t=100\,t_s$ and (b)  $t=500\,t_s$.  The arrows indicate planar projection of the director $\vec{u}$ and the colours indicate the phase order $\theta$ in these regions. In general there are two possible values  $\vec{u}$ and $\theta$ for the spinor at each simulation point [see Eq.~(\ref{EqDecomp})] because of the symmetry (\ref{EqDiscretesymm}), and we impose the further condition $u_y\ge0$.
We also show the locations of HQVs (see Sec.~\ref{SecHQVs}) with circulations $\sigma_1=1$ (black plus), $\sigma_1=-1$ (black triangle),  $\sigma_{-1}=1$ (white plus), and $\sigma_{-1}=-1$ (white triangle). Simulation parameters: $g_n=3g_s$, $q=-0.5\,q_0$, $n_c=10^4/\xi_s^2$, $l=200\,\xi_s$ and $N=256$ points.}
\end{figure}

\subsection{Late-time Universal coarsening dynamics}\label{SEC:Latetimeordering}
In addition to considering the emergence of local  spin-nematic order we wish to examine the spatial dependence of the textures (domains) that develop and how these evolve in time. In Fig.~\ref{microstatesfig} we visualize the system order in a region of a simulation soon after local order is established [Fig.~\ref{microstatesfig}(a)] and at a later time  [Fig.~\ref{microstatesfig}(b)]. This visualization is performed by decomposing the spinor field at each simulation point according to Eq.~(\ref{EqDecomp}) to obtain $\vec{u}(\mathbf{x})$ and $\theta(\mathbf{x})$. The results in Fig.~\ref{microstatesfig} demonstrate that the spin-nematic and superfluid (i.e.~global phase $\theta$) order tends to extend over larger length scales as time passes, showing that the system is coarsening toward an EP state with (quasi)-long range order.

To quantify the spatial dependence of the ordering we introduce the correlation functions 
\begin{align}
G_\phi(\mathbf{r},t)&=\frac{2}{n_c^2}\left\langle\mathrm{Tr}\left\{ Q(\mathbf{0})Q(\mathbf{r})\right\}\right\rangle_t,
\label{EqGorder1}\\
G_\theta(\mathbf{r},t)&=\frac{1}{n_c^2}  \left\langle \alpha_\perp^*(\mathbf{0})\alpha_\perp(\mathbf{r})\right\rangle_t,  
\label{EqG0order1}
\end{align}
for the spin-nematic and superfluid orders, respectively, evaluated at time $t$ after the quench. See Appendix \ref{AppCorrelfn} for more details about how these correlation functions relate to the atomic field operators.

To illustrate the use of these correlation functions, we consider the EP ground state spinor 
\begin{align}
\bm{\psi}_{\mathrm{EP}}=\sqrt{\frac{n_c}{2}}e^{i\theta}\left(\begin{array}{c}-e^{-i\phi} \\ 0 \\ e^{i\phi} \end{array}\right),\label{EqGSEP}
\end{align}
where the angle $\phi$ is associated with the spin-nematic order [i.e.~$\vec{u}\propto \cos\phi\,\hat{\mathbf{x}}+\sin\phi\,\hat{\mathbf{y}}$], and the global phase  $\theta$ is associated with the superfluid order. Taking  $\theta$ and $\phi$ to be spatially dependent random variables, we use  $\bm{\psi}_{\mathrm{EP}}$ to evaluate the correlation functions (\ref{EqGorder1}) and (\ref{EqG0order1}), yielding 
\begin{align}
G_\phi^{\mathrm{EP}}(\mathbf{r})&=\langle \cos 2[\phi(\mathbf{r})-\phi(\mathbf{0})]\rangle,\\  
G_\theta^{\mathrm{EP}}(\mathbf{r})&=\left\langle e^{i2[\theta(\mathbf{r})-\theta(\mathbf{0})]}\right\rangle.
\end{align}

In practice we compute the spin-nematic order parameter correlation function as
\begin{align}
G_\phi(r,t)=\!\int\!{d\Omega_r}{ }\!\int\!\frac{d^2\mathbf{x}'}{l^2}  \frac{2}{n_c^2}\langle \mathrm{Tr}\{Q(\mathbf{x}')Q(\mathbf{x}'+\mathbf{r})\}\rangle_t,\label{EqGorder}
\end{align} 
which includes averaging  to improve the statistics of our results:  $\langle\,\rangle_t$ denotes an average over trajectories (simulations with different seeding noise). The integral $\int d\Omega_r$ is an angular average in 2D position space (utilizing the isotropy of the system) and $l^{-2}\int d^2\mathbf{x}'$ denotes spatial averaging. The convolutions are efficiently computed using fast Fourier transforms. We also apply these additional averaging steps when computing the $G_\theta(r,t)$ correlation function.

\begin{figure*} 
 \includegraphics[width=7.0in]{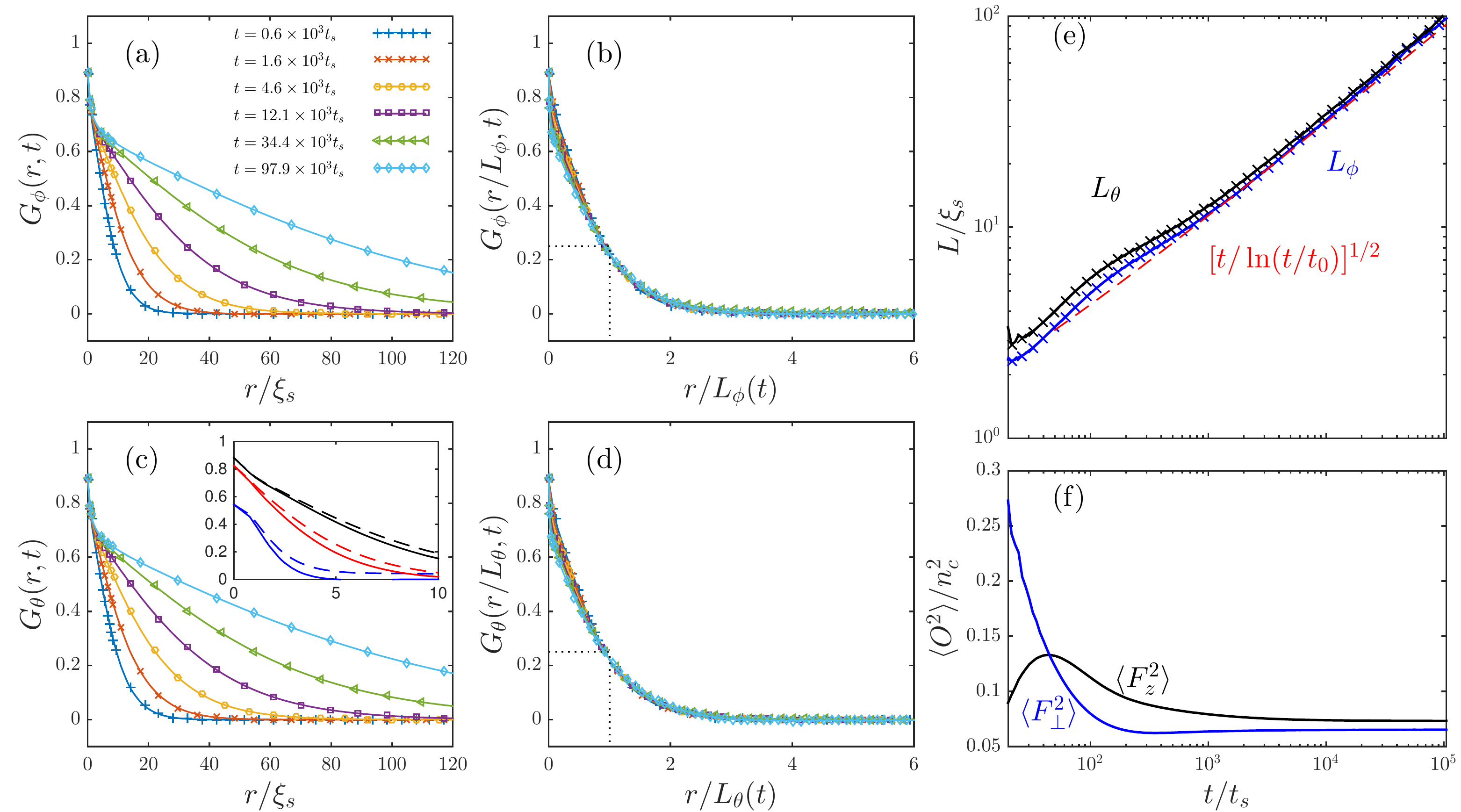} 
\caption{\label{scalingfig} Evolution and dynamic scaling of order parameter correlation functions. (a) The spin-nematic order correlation function $G_\phi$ at various times after the quench. (b) Collapse of the $G_\phi$ correlation functions when space is scaled by the length scale $L_\phi(t)$. (c) The superfluid order correlation function $G_\theta$ at various times after the quench. Inset compares $G_\phi$ (solid lines) and $G_\theta$ (dashed lines) at $t/t_s=25.6$ (blue), $99.0$ (red) and $399$ (black). (d) Collapse of the $G_\theta$ correlation functions when space is scaled by the length scale $L_\theta(t)$.  
(e) The evolution of the length scales $L_\phi$ and $L_\theta$ compared to a $[t/\ln(t/t_0))]^{1/2}$ growth law, where $t_0=0.5t_s$. Simulations are performed on domain of size $l=1600\,\xi_s$ covered by $N=2048$ points, and averaged over 15 trajectories. Interactions are $g_n=3g_s$, $n_c=10^4/\xi_s^2$ and $q=-0.5\,q_0$.}
\end{figure*}

Results for the evolution of $G_\phi(r,t)$ are shown in Fig.~\ref{scalingfig}(a). As time increases the correlation function is seen to decay more slowly, indicating that the in-plane spin-nematic order is extending over larger distances.  
We can investigate if the growth of this order exhibits dynamic scaling whereby the nematic domains are statistically self-similar at different times, up to an overall length scale that grows with time. This property often holds in the late time (when the domain sizes are much larger than microscopic length scales of the system) phase ordering dynamics of systems \cite{Bray1994}.
To verify dynamic scaling we demonstrate that the correlation function collapses to a universal (time-independent) function under time-dependent rescaling of space, i.e.~by showing that with an appropriate choice of $L_\phi(t)$ we have 
\begin{align}
H_\phi(r)=G_\phi(r/L_\phi(t),t).\label{Hphi}
\end{align}
Results showing the collapse are presented in Fig.~\ref{scalingfig}(b), where we have taken $L_\phi(t)$ to be the correlation length defined by the distance over which the correlation function decays to $\frac{1}{4}$ of its local value, i.e.~$G_\phi(L_\phi,t)=\frac{1}{4}G_\phi(0,t)$. The collapse is reasonably good except at short length scales ($r\ll L_\phi$) where the correlation function sharpens as $t$ increases. 

The length scale $L_\phi(t)$ is not unique and can be multiplied by a constant and still yield correlation function collapse. However, as chosen $L_\phi(t)$ gives a reasonable characterization of the domain size\footnote{Domain size cannot be uniquely defined because the in-plane nematic order varies continuously.} in the ordering EP system.
From considering the evolution of $L_\phi(t)$ we can extract the dynamic critical exponent $z_\phi$  as $L_\phi(t)\sim t^{1/z_\phi}$, providing a key characterization of dynamic universality class of the system.
In Fig.~\ref{scalingfig}(e) we show the time evolution of $L_\phi(t)$ on a log-log graph and find that at late times ($t\gtrsim10^3\,t_s$) this grows as $L_\phi(t)\sim [t/\ln(t/t_0)]^{1/2}$, i.e.~with a dynamic critical exponent of $z_\phi=2$  and logarithmic corrections. We find that the growth law exhibits a slight bulge (i.e.~above the asymptotic growth law) extending from early times  up until times of the order $10^3t_s$. We find that this correlates with the time period over which the magnetic fluctuations evolve appreciably in the system [see  Fig.~\ref{scalingfig}(f)], suggesting that the decay of magnetic fluctuations may set an important time scale for the system entering into the late-time coarsening regime (also see \cite{Kang2017a}).

The $L_\phi(t)\sim [t/\ln(t/t_0)]^{1/2}$ growth law we obtain here is the same form of growth known from the dissipative 2D XY model \cite{Yurke1993a,Bray2000a} (also see \cite{Kulczykowski2017a}), and was established in early work considering the coarsening dynamics of smectic liquid crystal films \cite{Pargellis1992a} (also see \cite{Nam2011,Singh2012a,Singh2013a}). Singh \textit{et al.}~\cite{Singh2012a} have predicted an analytic form of $H_\phi$ for nematic liquid crystals, which they have favourably compared to the results of Monte Carlo simulations using of a spin-nematic liquid crystal model \cite{Blundell1992a}. We however, find that this result is not a good fit to the $H_\phi$ we obtain.

We can also consider the superfluid scaling in this system, with examples of the evolving $G_\theta$ correlation function shown in Fig.~\ref{scalingfig}(c). We verify dynamic scaling in a similar way to the spin-nematic order by finding a length scale  $L_\theta(t)$ such that we have correlation function collapse:
\begin{align}
H_\theta(r)=G_\theta(r/L_\theta(t),t).
\end{align}
Results showing this collapse are presented in Fig.~\ref{scalingfig}(d), where again we have taken $L_\theta(t)$ to be  the distance over which the correlation function decays to $\frac{1}{4}$ of the its local value. These results also reveal that the late-time superfluid correlation function $G_\theta$ has a similar shape to the spin-nematic correlation function $G_\phi$. By definition both correlation functions have the same local value, i.e.~$G_\theta(0)=G_\phi(0)=\langle \mathcal{A}_\perp^2\rangle/n_c^2$. However, in general the superfluid correlation function decays more slowly and has a slightly longer characteristic length than the spin-nematic correlation function [e.g.~see inset in Fig.~\ref{scalingfig}(c)].

 In Fig.~\ref{scalingfig}(e) see that $L_\theta$ grows in a similar way to $L_\phi$, consistent with the same dynamical critical exponent, i.e.~$z_\theta\approx z_\phi\approx2$ (to within $\log$-corrections). Thus we find that the superfluid and spin-nematic order grow together in this system. This is different to recent results for the ordering of an EA ferromagnetic phase of a spin-1 condensate, which found that the superfluid order grows significantly slower than the spin order \cite{Andreane2017a} (also see \cite{Williamson2017a}).

\begin{figure} 
 \includegraphics[width=3.40in]{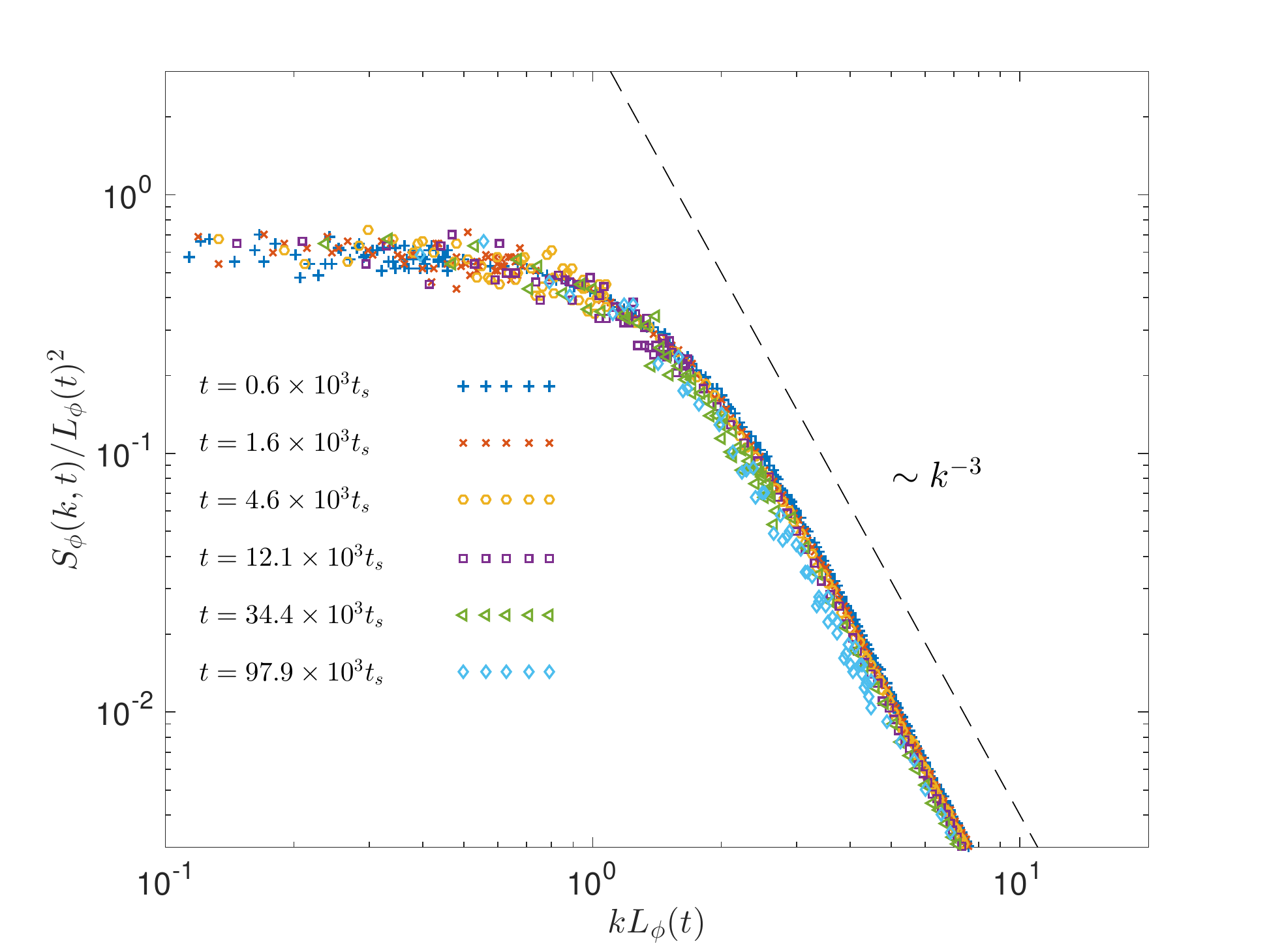} 
\caption{\label{Skfig} $S_\phi$ structure factor scaled by $L_\phi(t)$ to reveal scaling collapse. The power law decay for $kL_\phi>1$ reveals the Porod tail, with a guide line indicating $k^{-3}$ scaling for reference. Other parameters as in Fig.~\ref{scalingfig}.}
\end{figure}

It is conventional to also analyze the structure factors associated with the order parameter correlation function. The structure factor for spin-nematic order is defined as
\begin{align}
S_\phi(\mathbf{k},t)=\int d^2\mathbf{r}\,G_\phi(\mathbf{r},t)e^{i\mathbf{k}\cdot\mathbf{r}}.
\end{align}
The structure factors also collapse with dynamic scaling according to
\begin{align}
S_\phi(\mathbf{k},t)=L_\phi(t)^2\hat{h}\left(\mathbf{k}L_\phi(t)\right),
\end{align}
where $\hat{h}$ is the Fourier transform of $H_\phi$ (\ref{Hphi}). Results for the $S_\phi$ structure factor are shown in Fig.~\ref{Skfig}. For $k$ vectors in the range $L_\phi^{-1}<k\ll\xi_s^{-1}$  (i.e.~length scales between the microscopic healing length and the domain size) the structure factor exhibits a power law decay
 that is approximately of the form $k^{-3}$. This differs from the generalized Porod law result of $k^{-4}$ decay expected in 2D spin models \cite{Bray1991a,Singh2013a}. The $k^{-3}$ decay law is also found for the first order structure factors (single-particle momentum spectra) in studies of binary condensates in relevant regimes \cite{Karl2013a}, and is analyzed in terms of turbulence scaling.

We can  similarly define a superfluid structure factor $S_\theta$ from  $G_\theta$. This structure factor has a similar collapse and power-law decay to what we have presented for $S_\phi(k)$.

\subsection{Topological defects}\label{SecHQVs}

\begin{figure}
\includegraphics[width=3.0in]{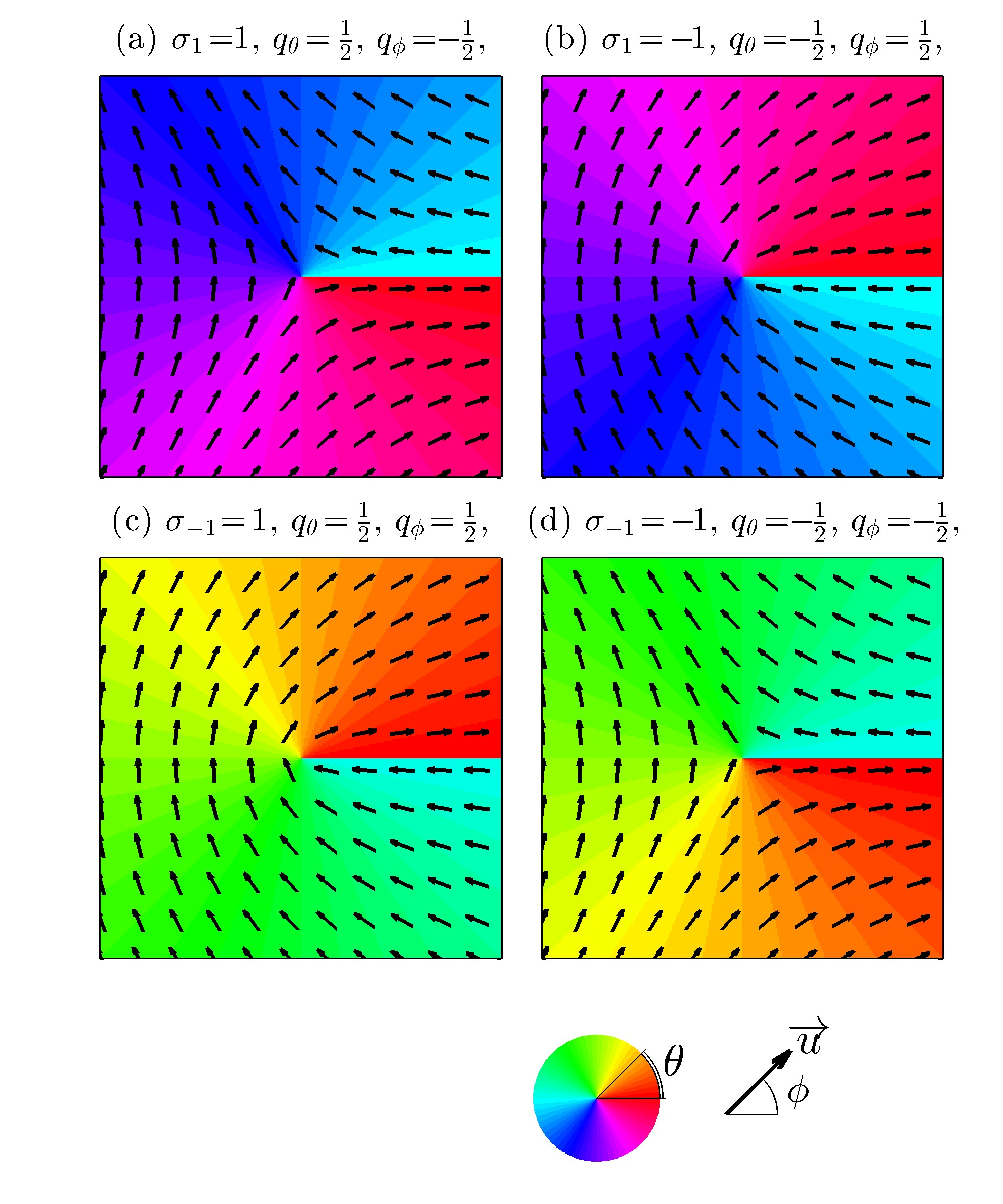}
\caption{\label{HQVs} Four types of HQVs that can occur in the EP phase are illustrated in subplots (a)-(d)  labelled by their winding numbers. }
\end{figure}

It is of interest to consider HQVs, which are the topological defects supported by the EP order parameter. To illustrate the properties of HQVs we first consider a single HQV located at the origin. Away from the core the wave function is approximately of the form  
\begin{align}
\bm{\psi}_{\mathrm{vort}}&=\sqrt{\frac{n_c}{2}}e^{iq_\theta\varphi}\left(
\begin{array}{c}
-e^{-iq_\phi\varphi}\\
0\\
e^{iq_\phi\varphi}
\end{array}
\right)\label{EqQform}
\sim
\left(
\begin{array}{c}
-e^{-i\sigma_1\varphi}\\
0\\
e^{i\sigma_{-1}\varphi}
\end{array}
\right), 
\end{align}
where we have set $\theta\to q_\theta\varphi$ and $\phi\to q_\phi\varphi$ in Eq.~(\ref{EqGSEP}),   $\varphi$ is the azimuthal angle about the core, and $\{q_\theta,q_\phi\}$ are the winding numbers. 
In Eq.~(\ref{EqQform}) we have also introduced the component windings
\begin{align}
\sigma_{\pm1}\equiv q_\theta\mp q_\phi,
\end{align}
where $\sigma_m$ denotes the net phase winding in the $m$-th component of the field.
The $\sigma_m$ must be integer for the field to be single valued.
The cases $\sigma_1=\pm1$ (with $\sigma_{-1}=0$) and $\sigma_{-1}=\pm1$ (with $\sigma_{1}=0$) define the four HQVs, corresponding to $q_\theta=\pm\frac{1}{2}$,  $q_\phi=\pm\frac{1}{2}$, i.e.~vortices with half-quantized values of the windings in $\theta$ and $\phi$ (see Fig.~\ref{HQVs}).  
 
Much of our theoretical understanding of HQV dynamics has come from studies of miscible two-component condensates \cite{Ohberg2002a,Eto2011a,Kasamatsu2016a}, which also support HQVs (also see \cite{Ji2008a}).
Notably, Eto \textit{et al.~}\cite{Eto2011a} have shown that the interaction potential between two HQVs separated by a distance $R$ (for $R\gg\xi_s$) is of the form
\begin{align}
U_{\mathrm{int}}\,\propto\,\kappa \ln R,
\end{align}
where  
\begin{align}
\kappa=q_{\theta}^{(1)}q_{\theta}^{(2)}+q_{\phi}^{(1)}q_{\phi}^{(2)}=\frac{1}{2}\sum_{m=\pm1}\delta_{\sigma_m^{(1)},\sigma_m^{(2)}}, 
\end{align}
with $(q_{\theta}^{(1)},q_{\phi}^{(1)})$ and $(q_{\theta}^{(2)},q_{\phi}^{(2)})$, [or $(\sigma_1^{(1)},\sigma_{-1}^{(1)})$ and $(\sigma_1^{(2)},\sigma_{-1}^{(2)})$] being the sets of winding numbers specifying  HQV 1 and 2, respectively. For the case where both HQVs have winding in the same component (i.e.~both having $|\sigma_1|=1$ or  $|\sigma_{-1}|=1$ ) then $|\kappa|=\frac{1}{2}$ and the interaction is of the same form as that for $U(1)$ vortices in a scalar condensate. When the vortices occur in different components $\kappa=0$ and there is no long ranged interaction. However, a short ranged repulsive interaction is predicted, extending over a length scale comparable to  the vortex core size \cite{Ji2008a,Eto2011a,Shirley2014a}. Two HQVs with opposite circulation in the same component (e.g.~a HQV with $\sigma_1=1$ and a HQV with $\sigma_1=-1$) can collide and annihilate, as has been recently observed in experiments  \cite{Seo2016a}.

 \begin{figure}
\includegraphics[width=3.2in]{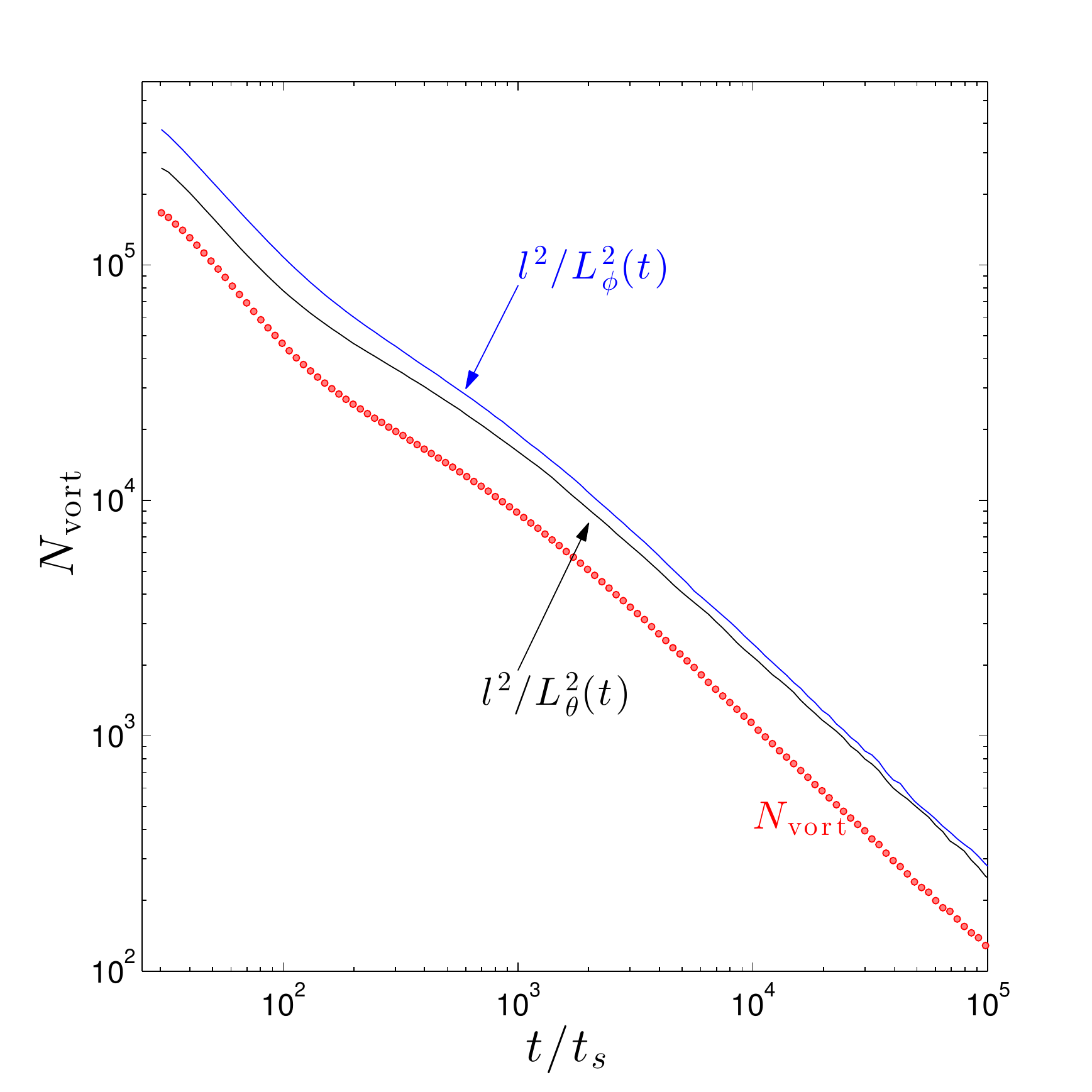}
\caption{\label{vortexfig} HQV  number as a function of time for the simulation case examined in Fig.~\ref{scalingfig}. The vortex number is computed as the total number of unit phase winding singularities in the $m=\pm1$ components averaged over the trajectories. The vortex number is compared to the \textit{number of domains} $l^2/L_\nu^2$, using the characteristic length scales $L_\nu=\{L_\phi,L_\theta\}$ [from Fig.~\ref{scalingfig}(e)] as labelled in the plot.}
\end{figure}

Coarsening dynamics can be viewed in terms of the dynamics of topological defects of the order parameter which are generated in the early stages of the quench dynamics. The windings associated with these defects disrupt the order, and as they mutually annihilate order is able to extend over larger length scales. We show the locations of HQVs in Fig.~\ref{microstatesfig}, which reveals a qualitative relationship between the domain sizes and  the vortex locations. To quantify the role of defects we detect the number of vortices in our simulations during the evolution. In practice we count the total number of integer phase windings in the $m=\pm1$ components. In the early time dynamics not all vortices detected are HQVs, but  we find that only HQVs persist at late times $(t\gtrsim100\,t_s$). In Fig.~\ref{vortexfig} we show the averaged total number of vortices $N_{\mathrm{vort}}$ as a function of time. The number of vortices decreases as the coarsening progresses. We can compare these results to the characteristic length scales discussed in Sec.~\ref{SEC:Latetimeordering}. Crudely, if the characteristic length scale is taken to be the distance between vortices then we would expect
\begin{align}
N_{\mathrm{vort}}(t)\sim\frac{l^2}{L_\nu(t)^2},\qquad\nu \in \{\phi, \theta\}.\label{EqNvort}
\end{align}
We have added these results for the characteristic length to Fig.~\ref{vortexfig} verifying that the relationship (\ref{EqNvort}) holds.

As we noted above a pair of $\sigma_{1}=1$ and $\sigma_{1}=-1$ HQVs (or a $\sigma_{-1}=1 $ and $\sigma_{-1}=-1$ pair) evolve similarly to a vortex anti-vortex pair in a  scalar condensate, and have the potential to mutually annihilate. In such a case each component vortex experiences a Magnus force which causes the pair to move with uniform velocity in a direction perpendicular to the line joining them. Such motion, without some other source of dissipation, does not lead to the vortices meeting and annihilating. This is in contrast to oppositely charged polar core spin vortices (the topological defects of  the easy-plane ferromagnetic phase) that accelerate towards each other and annihilate \cite{Turner2009,Williamson2016c}. 
We expect that in our system dissipation will arise from the interaction between the vortices and the sound waves (spin waves) excited by the quench.  
However, recent results on HQVs suggest an additional dissipative mechanism even in the absence of spin waves:  GPE simulations of a quiet binary condensate (without excitation) have found that such a pair of HQVs move together and annihilate (see Sec.~IV of \cite{Kasamatsu2016a}).  This effect was observed to be dependent on the interaction parameter regime, only occurring for $\gamma>0.5$, where $\gamma$ is the ratio of the inter- to intra-species interaction in the binary condensate. In the spin-1 system\footnote{This mapping is made by neglecting the $\psi_0$ component in the spin-1 GPE.} this parameter relates to the interaction parameters as $\gamma\approx(g_n-g_s)/(g_n+g_s)$.  Since our main simulations presented are for $\gamma=0.5$, where this additional dissipative effect is expected to be negligible, it is of interest to see if our coarsening dynamics changes for a larger value of $\gamma$. To explore this issue we have conducted quench simulations for $g_s=g_n/12$ ($\gamma\approx0.85$). The results for these simulations are roughly comparable to our main results in Fig.~\ref{scalingfig} (which are for $g_s=g_n/3$), and do not indicate that the coarsening proceeds at a faster rate.  Nevertheless a better understanding of HQV dynamics, particularly in the spin-1 system at finite $q$ values, would be a valuable direction for future research.   Also, a more detailed study of the dynamics and correlations between HQVs during the coarsening will be needed to illuminate the microscopic processes that are important in the system evolution (c.f.~\cite{Schole2012a}).

\section{Conclusion}\label{Sec:Conclude}
In this paper we have presented a theory for quantifying order formation in an anti-ferromagnetic spin-1 condensate. We have used this to study the dynamics of a quasi-2D system quenched into the EP spin-nematic phase. This topic has been of growing interest with a number of experimental developments motivating this work. This includes studies of correlations and spatial ordering in a quasi-one-dimensional system \cite{Bookjans2011b,Vinit2013a}, and evolution of magnetic fluctuations and HQV formation in a quasi-2D system \cite{Kang2017a}. A key issue has been identifying appropriate observables to quantify spin-nematic order.  This issue has been explored by Zibold \textit{et al.}~\cite{Zibold2016a} who developed a novel measurement scheme to demonstrate  spin-nematic order in   the single mode regime \cite{Zibold2016a}. 
We motivate and define order parameters for the system to quantify the spin-nematic and superfluid order, and in doing this we have connected our formalism to quantities that have already been measured in experiments. 

We have also studied the universal coarsening regime emerges at late-times after the quench. We evaluate the evolution of the order parameter correlation functions by averaging over an ensemble of large-scale numerical simulations and show that both types of order exhibit dynamic scaling, with a  characteristic length scale that grows as  $L\sim [t/\log (t/t_0)]^{1/2}$.  Our results also show that the coarsening is determined by the mutual annihilation of HQVs produced in the early stages of the quench. 
In experiments it may be difficult to directly measure the order parameter correlation function, whereas the average distance between HQVs (which can be directly imaged  \cite{Seo2015a,Seo2016a,Kang2017a}) will be a more convenient method to measure a characteristic length scale of order in the system. 
 
Having developed and applied formalism for non-ferromagnetic ordering in a spin-1 system we open the door to other studies of ordering in spinor systems. This includes the rich array of spin order that emerge in higher spin systems (e.g.~see \cite{Kawaguchi2012R}).

\section*{Acknowledgments}
The authors acknowledge support from the Marsden Fund of the Royal Society of New Zealand.  PBB thanks Y.~Kawaguchi for feedback on an early draft of the formalism,   acknowledges useful discussions with L.~Williamson, and  thanks B.T.~Wong for support of this research.
 
\appendix
\section{Planar treatment of spin-nematic order}\label{AppPlanar}
 We can formulate our order parameters by considering the cartesian spinor field projected onto the plane:
\begin{align}
\vec{\psi}_\perp\equiv(\psi_x,\psi_y)^T.
\end{align} 
Recalling $\psi_x=\frac{1}{\sqrt{2}}(\psi_{-1}-\psi_{1})$,   $\psi_y=-\frac{i}{\sqrt{2}}(\psi_1+\psi_{-1})$, we see that the planar treatment only depends on the $\{\psi_1,\psi_{-1}\}$ spherical components of the spinor.

We now proceed to develop a mathematical description of the spin properties of the planar-spin system analogously to the three-dimensional treatment developed in Sec.~\ref{SecPDSymmetries}.
We can decompose the planar spinor into two real planar vectors
\begin{align}
\vec{\psi}_\perp=e^{i\theta_\perp}(\vec{u}_\perp+i\vec{v}_\perp),\label{decompperp}
\end{align}
which are orthogonal and satisfy the normalization condition 
\begin{align}
|\vec{u}_\perp|^2+|\vec{v}_\perp|^2=n_\perp, \label{uvnorm}
\end{align}
where $n_\perp=\vec{\psi}_\perp^*\cdot\vec{\psi}_\perp=n_1+n_{-1}$.
We choose $\vec{u}_\perp$ to be the effective planar director and take it to be the longest vector, i.e.~$|\vec{u}_\perp|^2\ge\frac{1}{2}n_\perp\ge|\vec{v}_\perp|^2$.
We emphasize that the vectors $\{\vec{u}_\perp,\vec{v}_\perp\}$ are not in general the projected versions of the three-dimensional vectors in Eq.~(\ref{EqDecomp}) (e.g.~projection of $\{\vec{u},\vec{v}\}$ does not preserve their orthogonality).

Because our vectors are 2D we can only obtain a $z$-component of the cross product, which yields the usual $F_z=|\psi_1|^2-|\psi_{-1}|^2$ magnetization density, i.e.
\begin{align}
F_z=-i\vec{\psi}_\perp^*\times\vec{\psi}_\perp=2\vec{u}_\perp\times\vec{v}_\perp.
\end{align}
The $m=0$ component projected out of the spinor prohibits us from quantifying the transverse magnetization. 
The singlet-amplitude to the planar system is defined as
\begin{align}
\alpha_\perp\equiv\vec{\psi}_\perp\cdot\vec{\psi}_\perp=-2\psi_1\psi_{-1}\label{alphaperpapp}
\end{align}
and we have the relation [c.f.~Eq.~(\ref{EqAlignReln})]
\begin{align}
F_z^2+|\alpha_\perp|^2=n_\perp^2.
\end{align}

We can construct a symmetric traceless tensor  [i.e.~the one introduced in Eq.~(\ref{EqQ})] as
\begin{align}
Q&\equiv\frac{n_\perp}{2}I_2- \frac{1}{2}\left(\vec{\psi}_\perp^*\otimes\vec{\psi}_\perp +\vec{\psi}_\perp\otimes\vec{\psi}_\perp^* \right),\\ 
&=\frac{n_\perp}{2}I_2-  \left(\vec{u}_\perp\otimes\vec{u}_\perp +\vec{v}_\perp\otimes\vec{v}_\perp \right).\label{Quv}
\end{align}
As noted in Sec.~\ref{SecOrderEAEP}  the elements of $Q$ in spherical spinor components are $Q_{xx}=\mathrm{Re}\{\psi_{1}^*\psi_{-1}\}=-Q_{yy}$ and $Q_{xy}=\mathrm{Im}\{\psi_{1}^*\psi_{-1}\}$, with $\mathrm{det}(Q)=-n_1n_{-1}$.

By   inspection of Eq.~(\ref{Quv}) we see $\{\vec{u}_\perp,\vec{v}_\perp\}$ are eigenvectors of $Q$ with eigenvalues  $\lambda_u=\frac{1}{2}{n_\perp}-|\vec{u}_\perp|^2$ and $\lambda_v=\frac{1}{2}{n_\perp}-|\vec{v}_\perp|^2$, respectively.
Given our convention to choose $\vec{u}_\perp$ as the longer vector we have that $\lambda_u$ is negative (i.e.~the director corresponds to the lowest eigenvalue). Because the matrix is traceless the eigenvalues are given by $\pm\sqrt{-\mathrm{det}(Q)}$, i.e.~$\lambda_u=-\sqrt{n_1n_{-1}}$ and $\lambda_v=\sqrt{n_1n_{-1}}$. The trace of $Q^2$ is then just the sum of the eigenvalues squared, and recalling  the transverse alignment $\mathcal{A}_\perp=|\alpha_\perp|=\sqrt{2n_1n_{-1}}$, we obtain
\begin{align}
\mathrm{Tr}(Q^2)=\frac{1}{2}\mathcal{A}_\perp^2.
\end{align}

We also note that   $Q$ can be written in the form
\begin{align}
Q&=\frac{\mathcal{A}_\perp}{2}\left(\begin{array}{cc}
\cos 2\varphi& \sin2\varphi \\
 \sin2\varphi & -\cos 2\varphi \\
\end{array}\right),
\end{align}
where we have introduced $\varphi\equiv\frac{1}{2} \mathrm{Arg}(\psi_1^*\psi_{-1})$, i.e.~$\psi_1^*\psi_{-1}=\frac{1}{2}\mathcal{A}_\perp e^{2i\varphi}$. Note that this has 
eigenvalues and eigenvectors:
\begin{align}
\lambda_u=-\frac{\mathcal{A}_\perp}{2},\quad \hat{\vec{u}}_\perp&=\left(
\begin{array}{c} \cos\varphi\\
\sin\varphi\end{array}
\right),\\
\lambda_v=+\frac{\mathcal{A}_\perp}{2},\quad \hat{\vec{v}}_\perp&=\left(
\begin{array}{c} -\sin\varphi\\
\cos\varphi\end{array}
\right),
\end{align}
where the hats emphasize that these are unit vectors.
We observe that the relative phase of the $\psi_1$ and $\psi_{-1}$ components directly determines the orientation $\varphi$ of the planar director $\vec{u}_\perp$. Note that this result is general for any spin-1 spinor, however for the particular case of the EP ground state (\ref{EqGSEP}) we have $\varphi\to\phi$, $\mathcal{A}_\perp\to n_c$.  

\section{Correlation functions}\label{AppCorrelfn}
Using the results of the previous section we can provide an alternative motivation for the correlation functions used in the paper.
Firstly, we will consider the orientation of the director at two different points in space. For a spin model this might be characterized by a correlation function of the form 
\begin{align}
G_u(\mathbf{r})=\langle |\hat{\vec{u}}(\mathbf{0})\cdot\hat{\vec{u}}(\mathbf{r})|^2\rangle=\frac{1}{2}\langle\cos(2[\varphi(\mathbf{0})-\varphi(\mathbf{r})]) + 1\rangle,
\end{align}
where the inner product is squared to account for $\vec{u}$ and $-\vec{u}$ being the same.
In terms of the fields our relevant quantity is the complex density  
$\Phi\equiv\psi_1^*\psi_{-1}=\frac{1}{2}\mathcal{A}_\perp e^{2i\varphi}$. Correlating this at two points in space we have   
\begin{align}
G_\Phi(\mathbf{r})&=\langle \Phi(\mathbf{0})\Phi^*(\mathbf{r})\rangle,\\
&=\langle \psi_1^*(\mathbf{0})\psi_{-1}(\mathbf{0})\psi_{-1}^*(\mathbf{r})\psi_1(\mathbf{r})\rangle,
\end{align} 
which is identical to $G_\phi$ as defined in (\ref{EqGorder1})  if we normalize by a factor of $4/n_c^{2}$ . 

From Eqs.~(\ref{decompperp}) and (\ref{alphaperpapp}) we see that the superfluid phase $\theta_\perp$ is related to the singlet-amplitude as 
\begin{align}
\alpha_\perp=-2\psi_1\psi_{-1}= -\mathcal{A}_\perp e^{2i\theta_\perp},
\end{align}
where we can take  $\theta_\perp=\frac{1}{2} \mathrm{Arg}(\psi_1\psi_{-1})$. Thus to correlate this superfluid order at two points we can consider the pairing-like field $\alpha_\perp$ at those two locations, i.e.~
\begin{align}
G_{\alpha_\perp}(\mathbf{r})
&=\langle\alpha_\perp^*(\mathbf{0})\alpha_\perp(\mathbf{r})\rangle,\\
&=4\langle \psi_1^*(\mathbf{0})\psi_{-1}^*(\mathbf{0})\psi_{-1}(\mathbf{r})\psi_1(\mathbf{r})\rangle.
\end{align} 
Normalizing by a factor of $n_c^{-2}$ gives $G_\theta$  [Eq.~(\ref{EqG0order1})].

\end{document}